\documentclass[conference]{IEEEtran}
\IEEEoverridecommandlockouts
\usepackage{cite}
\usepackage{amsmath,amssymb,amsfonts}
\usepackage{algorithmic}
\usepackage{graphicx}
\usepackage{textcomp}
\usepackage{xcolor}
\usepackage{tabularray}
\usepackage{hyperref}

\usepackage{booktabs, makecell, tabularx}
\newcolumntype{C}{>{\centering\arraybackslash}X} % centered version of "X" type

\usepackage{stfloats}
\usepackage{siunitx}
\usepackage{caption}

\usepackage{lipsum}

\usepackage{enumitem}
\setlist[itemize]{align=parleft,left=0pt..1em}

\def\BibTeX{{\rm B\kern-.05em{\sc i\kern-.025em b}\kern-.08em
    T\kern-.1667em\lower.7ex\hbox{E}\kern-.125emX}}
\begin{document}

\title{VoiceBank-2023: A Multi-Speaker Mandarin Speech Corpus for Constructing Personalized TTS Systems for the Speech Impaired\\

\thanks{This work was primarily supported by the NSTC of Taiwan under Contract No. 109-2221-E-305-010-MY3, and the Taiwan Motor Neuron Disease Association under Contract No. 111A617201. This work was also supported by the NSTC of Taiwan under Contract No. 112-2425-H-305 -002, National Taipei University, Taiwan under Contract No. 112-NTPU-ORDA-F-004, and Shih Chien University, Taiwan under Contract No. USC-111-03-01020}
}

\author{\IEEEauthorblockN{\textsuperscript{1}Jia-Jyu SU, \textsuperscript{1}Pang-Chen LIAO, \textsuperscript{1}Yen-Ting LIN, \textsuperscript{1}Wu-Hao LI, \textsuperscript{2}Guan-Ting LIOU, \textsuperscript{3}Cheng-Che KAO,\\\textsuperscript{3}Wei-Cheng CHEN,\textsuperscript{2}Jen-Chieh CHIANG, \textsuperscript{3}Wen-Yang CHANG, \textsuperscript{1}Pin-Han LIN, \textsuperscript{1}Chen-Yu CHIANG}
\IEEEauthorblockA{\textit{\textsuperscript{1}National Taipei University, Taiwan,
\textsuperscript{2}National Yang Ming Chiao Tung University, Taiwan,
\textsuperscript{3}AcoustInTek Co., Ltd., Taiwan}\\
boss.su2000@gmail.com, s410986030@gm.ntpu.edu.tw, \{d26923050, hank12451, tn00320663\}@gmail.com,\\ \{erickao, weichengchen, jackchiang, chrischang\}@acoustintek.com, joseph861030@gmail.com, cychiang@mail.ntpu.edu.tw}
}
\maketitle

\begin{abstract}
Services of personalized TTS systems for the Mandarin-speaking speech impaired are rarely mentioned. Taiwan started the VoiceBanking project in 2020, aiming to build a complete set of services to deliver personalized Mandarin TTS systems to amyotrophic lateral sclerosis patients. This paper reports the corpus design, corpus recording, data purging and correction for the corpus, and evaluations of the developed personalized TTS systems, for the VoiceBanking project. The developed corpus is named after the VoiceBank-2023 speech corpus because of its release year. The corpus contains 29.78 hours of utterances with prompts of short paragraphs and common phrases spoken by 111 native Mandarin speakers. The corpus is labeled with information about gender, degree of speech impairment, types of users, transcription, SNRs, and speaking rates. The VoiceBank-2023 is available by request for non-commercial use and welcomes all parties to join the VoiceBanking project to improve the services for the speech impaired.
\end{abstract}


\section{Introduction}
The voice of each individual may be regarded as his/her identity. Amyotrophic lateral sclerosis (ALS) patients will gradually lose the ability to control their muscles, which affects the control of the glottal fold and shape of the vocal tract, and become difficult to pronounce and communicate smoothly. ALS patients are encouraged to record their voices before they become dysarthria. The recorded speech can be used to construct personalized text-to-speech (TTS) systems, which serve as speech-generating devices (SGD) for augmentative and alternative communication (AAC).

In English-speaking countries, many companies or research institutes are providing services to make personalized TTS systems for ALS patients to use. The significant one is Model Talker \cite{ModelTalker}, which is the earliest and largest research platform in the US, established by the Nemours Speech Research Laboratory located at the Alfred I. duPont Hospital for Children in Delaware, US. With the advances in speech technologies, the following commercial SGD providers can be found: Cereproc Cerevoice ME \cite{CereProc},  VocalID \cite{VOCALiD}, Acapela my-own-voice DNN \cite{Acapela}, the Voice Keeper \cite{VoiceKeeper}, and SpeakUnique\footnote{SpeakUnique powers the user-friendly voice-banking platform, “I Will Always Be Me” (https://www.iwillalwaysbeme.com/)} \cite{SpeakUnique}.

In 2011, the Motor Neurone Disease (MND) Association in the UK \cite{mndauk} started the voice banking project to recommend that patients deposit their voices and messages with the services provided by the abovementioned institutes or companies. Similarly, in 2018, the ALS association in the US initiated Project Revoice \cite{usrv} to encourage patients to contact the institutes or vendors mentioned above to build customized TTS systems for themselves before they become dysarthria.

Though the AI boom in recent years has made the performances of TTS systems much better, services and techniques to build personalized TTS systems for Mandarin-speaking ALS patients are rarely mentioned. The main reasons are that the speech data that ALS patients could provide is extremely scarce and that the personalized TTS techniques make it hard to build acceptable and usable services for patients with such little data. Inspired by the Project Revoice raised in the US, Taiwan started the VoiceBanking project \cite{VoiceBankProject} in 2020, aiming to build a complete set of services to deliver personalized Mandarin TTS systems to ALS patients. Enrolled ALS patients are encouraged to record (or bank) their voices before dysarthria occurs and to utilize constructed personalized TTS systems as SGDs when losing the ability to speak.

This paper reports the current progress (2020-2023) of the design of the corpus, recording of the corpus, data purging/correction for the corpus, and the evaluations of the developed personalized TTS systems. We name the developed corpus the VoiceBank-2023 speech corpus because the database was released in 2023. We summarize the specification for the VoiceBank-2023 corpus in TABLE~\ref{table:Brief_specification_for_the_VoiceBank2023_corpus}.

\begin{table*}[t]
  \caption{Specification for the VoiceBank-2023 corpus.}
  \centering
  \resizebox{\textwidth}{!}{\begin{tabular}{|l|l|}
    \hline
    Corpus Name & VoiceBank-2023 (URL: \href{https://github.com/VoiceBank-NTPU-TW/VoiceBank-2023}{https://github.com/VoiceBank-NTPU-TW/VoiceBank-2023}\\ \hline
        Language & mostly Taiwanese Mandarin \\ \hline
        
        Text/Prompt materials & 1) Part-1 (VoiceBanking): 133 short paragraphs, and 2) Part-2 (Common Phrases): 556 common phrases \\ \hline

        Speaking Style & 1) read speech for  Part-1 (VoiceBanking), and 2) spontaneous like for Part-2 (Common Phrases) \\ \hline
        
        Uses & 1) personalized TTS, 2) assessments of dysarthria, voice quality (jitter/shimmer), and sound quality (regarding recording) \\ \hline

        \makecell[l]{\# of Speakers (speaker types, gender\\~~~~~~~~~~~~~ , dysarthria degree)} & {\makecell[l]{111(all) = 39(ALS patients) + 63(voice donors) + 9(unknowns) = 47(female) + 64(male)\\= 86(degree 1: high speech intelligibility) + 11(degree 2) + 12(degree 3), 2(degree 4: low speech intelligibility)}}\\ \hline
                
        \makecell[l]{\# of Utterances (prompt type, gender\\~~~, speaker type, dysarthria degree)} & \makecell[l]{12,875(all) = 7,625(Part-1, VoiceBanking) + 5,250(Part-2, Common Phrases) = 5,677(female) + 7,198(male)\\= 8,876(patient) + 3,875(donor) + 124(unknown) = 8,760/2,246/1,849/20(degree 1/2/3/4)} \\ \hline
                 
        Total Duration (hours)& \makecell[l]{29.78(all) =  28.18(Part-1:VoiceBanking) + 1.60(Part-2: Common Phrase) = 12.47(female) + 17.31(male)\\ = 17.66(patients) + 11.78(donors) + 0.34(unknowns) = 19.37/5.74/4.58/0.09(degree 1/2/3/4)}\\ \hline

        \makecell[l]{Duration for \\each Speaker (minutes)}& Part-1 (VoiceBanking): 15.37$\pm$10.97, Part-2 (Common Phrases): 5.99$\pm$5.34\\ \hline

        \# of Syllables & \makecell[l]{360,586(all) = 342,486(Part-1:VoiceBanking) + 18,100(Part-2: Common Phrase) = 153,396(female) + 207,190(male)\\ = 185,401(patients) + 170,387(donors) + 4,798(unknowns) = 270,805/55,490/33,835/456(degree 1/2/3/4)}\\ \hline
        
        Utterance Length in Syllable & Part-1 (VoiceBanking): 44.13$\pm$9.03, Part-2 (Common Phrases): 3.30$\pm$0.54 utterance-wise mean$\pm$standard deviation\\ \hline
        
        Utterance Length in Second & Part-1 (VoiceBanking): 13.16$\pm$4.87, Part-2 (Common Phrases): 1.08$\pm$0.32 utterance-wise mean$\pm$standard deviation\\ \hline
        
        Waveform Encoding & linear PCM, 48kHz sample rate, 16-bit resolution, mono channel \\ \hline
        
        Microphone/Recording Environment & mostly USB quality microphone/mostly home or office \\ \hline

        Files for each Utterance & \makecell[l]{1) *.TextGrid: time alignments for phonetic (initial/final), syllabic (tone), and word (part of speech and punctuation marks), \\2) *.txt: raw text file in UTF-8, 3) *.wav: WAVE file}\\  \hline
  \end{tabular}}
  \label{table:Brief_specification_for_the_VoiceBank2023_corpus}
\end{table*}

This paper is organized as follows:
\begin{itemize}[leftmargin=*]
    \item Section II: Sharing the experience for designing a speech corpus of voice-banking for ALS patients.
    \item Section III: Illustrating the on-site recording and the design of an online web assessable voice-banking platform.
    \item Section IV: Illustrating how to make data purging when speakers and recording environment are not professional.
    \item Section V: Introduce the labeling and metadata of the VoiceBank-2023 corpus.
    \item Section VI: Reporting developed TTS systems with the VoiceBank-2023 corpus.
    \item Section VII: Conclusions and future works.

\end{itemize}

\section{Design of Corpus}

\subsection{Consideration}

With the fast development of AI technologies, it is easy to find open resources for constructing TTS systems. The famous open-source speech corpora for constructing TTS are AISHELL3 \cite{aishell-3} for Mandarin, VCTK \cite{vctk}, ARCTIC \cite{arctic}, libriTTS \cite{libtts}, and HiFiTTS \cite{hifi} for English. However, none of the corpora stated above are specially designed for constructing personalized systems for ALS patients. Since ALS patients gradually lose their ability to speak, it is desired to collect patients’ speech with the following considerations:

\subsubsection{Consistent Text Material} Because the corpus size for each patient is small, each personalized TTS system needs to be constructed by speaker adaptation technique. The VoiceBanking project used two large speech corpora, the Treebank-SR corpus \cite{Treebank} and the Danei corpus\footnote{The Danei speech corpus is a Mandarin-English mixed speech corpus owned by AcoustInTek Co., Ltd. and licensed for the VoiceBanking project} to train the reference/prior model. Each personalized TTS model is then jointly trained with the two large speech corpora. To make adaptation data consistent with the data for the reference model, the text material for the patients’ adaptation speech corpus is designed to be subsets of the Treebank-SR corpus.

\subsubsection{Phonetically-Balanced Voice-Banking With Limited Text Material} The most important purpose of speech recording for a patient is to let the TTS model \textbf{\textit{produce intelligible synthesized speech}} and \textbf{\textit{learn the speaker's identity}} from the recorded data. It is desired to collect speech data as small as possible to cover most pronunciations. Utterances for a speaker are recorded in a sorted sequence such that the utterances at the beginning incrementally cover most pronunciations quickly. 

\subsubsection{Common Phrases for Daily Life} To enrich the personalized TTS with communicative or expressive functions, recorded common phrases can be enrolled in the training of the TTS model or directly be played back when using AAC.

According to the considerations, the VoiceBank-2023 corpus was designed to have two parts with eight sub-corpora:
\begin{enumerate}[leftmargin=*]
    \item Part 1 - VoiceBanking (sub-corpora 1 and 2):
    \begin{itemize}
        \item Sub-corpus 1: covers all Mandarin initial and final types
        \item Sub-corpus 2: enlarge sample size for voice-banking
    \end{itemize}
    \item Part 2- Common Phrases (sub-corpora 3 to 8)
    \begin{itemize}
        \item Sub-corpora 3 to 8: comprised of 1 to $\geq 6$-character phrases to enrich the communicative functions
    \end{itemize}
\end{enumerate}

\subsection{Part-1: VoiceBanking (sub-corpora 1 and 2)}
\subsubsection{Source Text Corpus - the Treebank-SR Corpus}
The text materials for voice-banking are excerpted from the texts of the Treebank-SR corpus, which has been used in constructing Mandarin \cite{pg0} \cite{pg1} and Chinese dialects \cite{pg2} speaking-rate controlled text-to-speech system, and in studies about prosody grammar \cite{pg3}. In the VoiceBanking project, the Treebank-SR corpus is also used to construct the prior TTS model for speaker adaptation. The corpus comprises four parallel sub-corpora in fast, normal, medium, and slow speaking rates. Each sub-corpus has 376 utterances with 52,031 syllables. The average length of each utterance is 138.38 syllables, with a standard deviation of  24.97. Note that the text material for each utterance generally corresponds to a short paragraph. 

\subsubsection{The Problem of Source Text}
The lengths of the Treebank-SR corpus are too long for non-professional speakers to speak fluently. We thus need to sort the 376 text paragraphs with a phonetic coverage constraint and then segment these sorted texts into shorter sentences to ease recording. Note that we did not first delimit paragraphs into short sentences and then sort the short sentences because we tended to record speakers' speeches with more comprehensible text context. 

\subsubsection{Steps to Obtain Sub-corpora}
The prompt texts for the Part-1: VoiceBanking, i.e., $\{S_{j}\}|_{j=1,2,...,J}$ is derived from the 376 paragraphs, i.e., $\{P_{k}\}|_{k=1,2,...,376}$ by the following steps:

\textbf{Step 1: Calculating key values for sorting}:    For each $k$-th text paragraph of the 376 text paragraphs,i.e., $P_{k}$, we calculate: 1) $|P_k|$: number of syllables in the $k$-th paragraph, and 2)  $|T(P_k)|$: number of initial and final types that occur in the $k$-th paragraph; where $T(\cdot)$ is a unique operator. %{such that $T(P_k)=unique(I_{k,1}, F_{k,1},I_{k,2}, F_{k,2},...,I_{k,|P_k|}, F_{k,|P_k|})$; $I$ and $F$ denote initial and final, respectively; the subscript {$k,n$} represent $n$-th syllable in $k$-th paragraph.}
 
\textbf{Step 2: Initial paragraph sorting}: The 376 paragraphs are sorted by $|T(P_k)|$ and then by $|P_k|$. The sortings are in descending order. The 376 sorted text paragraphs are expressed by $\{P_{k}\}|_{k=1,2,...,376} \gets sort(\{P_{k}\}|_{k=1,2,...,376})$.
    
\textbf{Step 3: Obtaining the initial priority paragraph set}: Let $V=\{P_{k}\}|_{k=1}$ be the priority paragraph set, and let $k$ be the pivot paragraph index. The priority set $V$ is used to calculate the counts of acoustic units, i.e., initials and final, for checking if the sorted paragraphs from 1-st to the $k$-th paragraphs contain phonetically balanced acoustic units.

\textbf{Step 4: Updating the priority paragraph set and adjusting the sequential order of the paragraph set}: For $i=k$ to 376,
if $|T(\{V \cup P_i\})| > |T(V)|$ then update the sequential order of the paragraph set by swapping contents of $P_i$ and $P_k$, update the priority paragraph set by $V \gets V \cup P_k$, set $k=k+1$, and go to Step 5.

\textbf{Step 5: Checking criteria for terminating:} Let $C_{i}(V)$ denotes the count for the $i$-th initial or final type in the priority paragraph set $V$. If $k>376$ or $C_{i}(V) \geq 2$ for $i=1,2,...,I$, go to Step 6 or go to Step 4.

\textbf{Step 6: Obtaining the initial prompt texts for Part-1: VoiceBanking}: We manually segment the sorted paragraphs $\{P_{k}\}|_{k=1,2,...,376}$ into short sentences $\{S_{j}\}|_{j=1,2,...,J}$ in sequence. Note that each segmented short sentence is preferable to be semantically meaningful.

\textbf{Step 7: Embellishing prompt text for better readability}: Parts of the short sentences $\{S_{j}\}|_{j=1,2,...,J}$ are hard to read. We add a few words to these hard-to-read sentences to smooth the reading. These embellished short texts are regarded as the prompt texts for Part-1: VoiceBanking.

\begin{figure}[htbp]
\centerline{\includegraphics[width=1.0\linewidth]{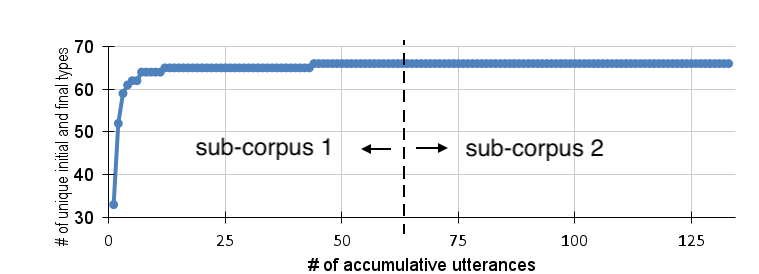}}
\caption{The number of unique acoustic units (initial/final) vs. accumulative utterances numbers.}
\label{fig:FIG1}
\end{figure}

Fig.~\ref{fig:FIG1} shows the number of unique acoustic units (initial/final) vs. accumulative utterances numbers. Note that Part-1 is delimited into two parts: sub-corpora 1 and 2. We encourage the enrolled patients to finish the recording for sub-corpus 1 because sub-corpus 1 covers almost all acoustic units. 

\subsection{Part-2: Common Phrases (sub-corpora 3 to 8)}

The text materials of 556 common phrases are excerpted from Appendix 14 of the textbook \cite{MotorSpeechDisorder}. These common phrases were collected as short lists to facilitate verbal communication for people with speech impairment. The phrases can be classified into six types by lengths in syllables, including lengths in 1, 2, 3, 4, 5, and $\geq 6$. TABLE~\ref{table:Statistics_of_the_common_phrases} shows the statistics of the common phrases collected in the VoiceBank-2023.

\begin{table}[!ht]
    \centering
    \caption{Statistics of the common phrases.}
    \label{table:Statistics_of_the_common_phrases}
    \begin{tabular}{lcccccc}
    \hline
        \textbf{sub-corpus index} & 3 & 4 & 5 & 6 & 7 & 8 \\ \hline
        \textbf{\# of utterance (phrase)} & 39 & 109 & 185 & 105 & 54 & 64 \\ \hline
        \textbf{phrase length in syllable} & 1 & 2 & 3 & 4 & 5 & $\geq 6$ \\ \hline
    \end{tabular}
\end{table}

\section{Corpus Recording}
Here, we report the recording process for the VoiceBank-2023 corpus from Jan-20 - June-23. The corpus recording is divided into four phases according to the purpose of research and recording environments. TABLE~\ref{table:brief_summarization_of_the_four_phrases} summarizes the four phrases. Phase 1 collected speech corpora by the onsite recording using the high-quality microphone with the help of audio technicians. On the other hand, in Phases 2-4, users banked their voices by logging into \href{https://voicebank.ce.ntpu.edu.tw/}{https://voicebank.ce.ntpu.edu.tw/}, which provides a self-service GUI for voice recording. 

\begin{table*}[t]
  \caption{A summary for the four phrases.}
  \centering
  \resizebox{\textwidth}{!}{\begin{tabular}{|l|l|l|l|l|}
    \hline
    ~ & \makecell{Phase 1} & \makecell{Phase 2} & \makecell{Phase 3} & \makecell{Phase 4} \\ \hline
        Description & onsite recording & \makecell{trial web-based service \\ for voice donor} & \makecell{trial web-based service\\for patient} & \makecell{web-based service\\announced} \\ \hline
        
        Dates & Apr. 2020 - Apr. 2021 & Jan.-Feb. 2022 & Feb. 2022 - Jan. 2023 & Mar.-Jun. 2023\\ \hline

        Purpose & \makecell[l]{Test the feasibility of the \\VoiceBanking sub-corpus (Part-1) in \\ constructing personalized TTS} & \makecell[l]{Test the feasibility of \\ the web-based recording\\at home by voice donors.} & \makecell[l]{Test the feasibility of \\ the web-based recording\\at home by ASL patient} & \makecell[l]{Test the feasibility of \\ the web-based recording\\at home for all} \\ \hline

        Microphone & audio-technica ATM73a & \multicolumn{3}{c|}{various USB microphones owned by speakers} \\ \hline

        Audio Interface & Steinberg UR RT-2 & \multicolumn{3}{c|}{USB microphones built-in} \\ \hline
        
        Audio Recording Software & Adobe Audition & \multicolumn{3}{c|}{login \href{https://voicebank.ce.ntpu.edu.tw/}{https://voicebank.ce.ntpu.edu.tw/} with a browser} \\ \hline
        
        Guidance for Recording & recording technicians & \multicolumn{3}{c|}{self service GUI} \\ \hline
        
        Prompts & printed on papers & \multicolumn{3}{c|}{displayed on GUI}\\ \hline
        
        Registration & manual & \multicolumn{2}{c|}{\makecell{manual in the backend of website}} & \makecell{online on website} \\ \hline
        \makecell[l]{Speakers \# (patient/donor/unknown)\\(gender) (dysarthria degree 1/2/3/4)} & \makecell[l]{20 (18/2/0)\\(6F/14M) (9/5/5/1)} & \makecell[l]{61 (0/61/0)\\(27F/34M) (61/0/0/0)} & \makecell[l]{18 (17/0/1)\\(7F/11M) (5/5/7/1)} & \makecell[l]{12 (4/0/8)\\(7F/5M) (11/1/0/0)}\\ \hline
  \end{tabular}}
  \label{table:brief_summarization_of_the_four_phrases}
\end{table*}

\subsection{Phase-1: Onsite recording during 2020/4 to 2021/4}

We provided on-site recording services for the enrolled patients to bank their voices at their homes with the help of the research team in person, ensuring that the recordings can be fully used in constructing personalized TTS systems. The on-site recording is adopted because of the following reasons:

\begin{enumerate}[leftmargin=*]
    \item This project needed to collect patients' voices as soon as possible before they become impaired.
    \item It is not easy for the enrolled patients to commute to the professional recording studio. The on-site recording may increase the patients' willingness to bank their voices.
    \item Patients would make more effort to read text materials than most people do. The on-site in-person assistance in recording may alleviate patients’ workload.
    \item  The on-site recording engineers ensured that the recording setups met high-quality requirements for constructing TTS. 
    \item Because Taiwan is densely populated, the on-site recording engineers help to reduce unwanted environmental interference from neighbors or outdoors.
\end{enumerate}

\subsection{Phases 2-4: Web-based recording (2022/1)}

Due to the COVID-19 pandemic becoming severe in Taiwan in 2021, the on-site recording was not feasible. Meanwhile, quality microphone headsets became prevalent because people needed to work from home with the conference call systems. We therefore developed the web-based recording platform - VoiceBank website (\href{https://voicebank.ce.ntpu.edu.tw/}{https://voicebank.ce.ntpu.edu.tw/}) to facilitate voice-banking for everyone. The functions of audio recording are provided by the RecordRTC library \cite{recordrtc}. The display and playback of a recorded waveform are powered by the wavesurfer.js library \cite{wsjs}. The recorded waveforms are in a format of linear PCM with a 48KHz sample rate and 16-bit resolution. Note that we turn off the functions of speech enhancement and auto-gain control of RecordRTC to avoid destroying the original speech quality. Instead, we encourage the enrolled speakers to use USB-terminal microphones to record their voices in a quiet environment.

\begin{figure}[htbp]
\centerline{\includegraphics[width=1.0\linewidth]{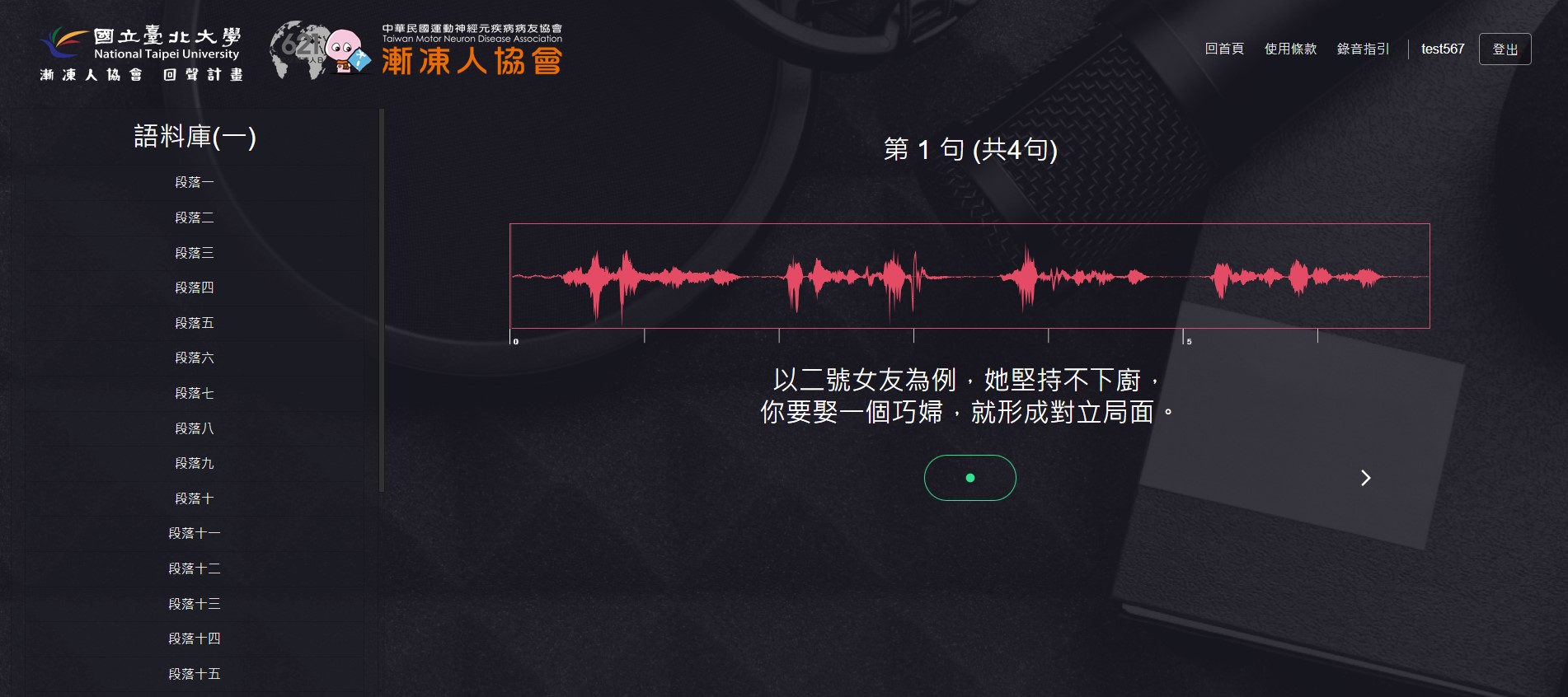}}
\caption{GUI of \href{https://voicebank.ce.ntpu.edu.tw/}{https://voicebank.ce.ntpu.edu.tw/}.}
\label{fig:FIG2}
\end{figure}

Fig.~\ref{fig:FIG2} shows the GUI for the recording page on the VoiceBank website. The left pane shows the selection for the sub-corpora and its including paragraphs. The right pane contains, from top to down, a sentence index displayer, a waveform displayer, a text prompt, and a recording control pane. To record the corpora, the users can first select a sub-corpus and one of its including paragraphs on the left pane. Once a paragraph is selected, the prompt of the first sentence of the paragraph of a sub-corpus will be shown. The waveform is shown on the waveform displayer if the sentence has been recorded or the waveform displayer shows a black background.

The user may push the green button on the recording control pane to start a take. When recording, the green button turns red, and the waveform displayer shows the recorded waveform in real time. The red button will turn green if a user presses the red button to stop recording. After stopping recording, the waveform recorded will be shown on the waveform displayer and saved in the backend of the VoiceBank website. Users may click on the waveform displayer to playback the recorded sentence. If a user is dissatisfied with the take, he/she can re-record the sentence. All the takes will be stored on the server, and only the last one will be shown to the user. The last take will be regarded as a usable sample.

\section{Data Purging and Correction for Phases 2-4}
\begin{figure}[htbp]
\centerline{\includegraphics[width=0.95\linewidth]{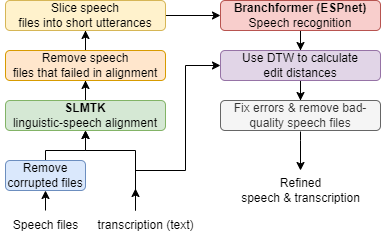}}
\caption{Flowchart of data purging and correction.}
\label{fig:FIG3}
\end{figure}

Since Phases 2 to 4 collected speech in a self-service fashion, the quality of recordings could not be as good as the ones recorded in Phase 1, where recording engineers supervised the on-site recording and corrected the transcription immediately after recording. Thus, we must conduct the data purging and correction process to remove unusable data or correct wrong transcription. Fig.~\ref{fig:FIG3} shows the flowchart of the process that can be roughly divided into three parts: file format checking, checking by force alignment, and checking by automatic speech recognition (ASR).

 First, the file format checking removed corrupted files with bad headers or zero lengths. Second, the checking by forced alignment removes the wave and text files that are not paired or cannot be aligned with each other. More specifically, we used the offline version of the SLMTK 1.0 \cite{slmtk} to conduct text analysis, acoustic feature extraction, and linguistic-speech alignment. The text analysis step obtained the sequences of linguistic units corresponding to the input text. The linguistic-speech alignment step produced TextGrid files containing various linguistic unit time information. By the linguistic-speech alignment step, we removed the wave and text pairs that do not generate TextGrid files. We found that the speech contents of the wave files that can not generate TextGrid files were often mismatched with the associated texts. 

Third, we use ASR to check the mismatch between wave and text files delicately. Specifically, we trained the \textit{Branchformer} \cite{bf} end-to-end ASR model on Aishell-1 corpus (85 hours) \cite{aishell-1} as a pre-trained model. This pre-trained model was then fine-tuned to the speech data comprised of the TCC-300 corpus \cite{tcc300} (25 hours) and the 6-hour VoiceBank data that passed the examination of the checking by forced alignment. The fine-tuning is needed because the VoiceBank and TCC-300 \cite{tcc300} corpora are in Taiwan-accent Mandarin while Aishell-1 is in mainland accent Mandarin. Because \textit{Branchformer} can not accept long speech input, the sentences recorded are segmented into several short utterances, and the \textit{Branchformer} took these short utterances as input. Pauses and/or punctuations delimited these short utterances according to TextGrid files obtained by checking by forced alignment.

Taking all the VoiceBank speech corpus as a test for ASR, character and sentence error rates were 4.0\% (among 300,570 target syllables) and 9.4\% (among 32,104 delimited short utterances), respectively. Because the character error rate is very low, we believe that the recognition result helps check the correctness of texts. We manually checked the short utterances of the character error rates higher than 30\%. The utterances with high error rates were fixed with correct transcriptions or discarded if the utterances were very hard to correct quickly.

\section{Labeling and Metadata}
Each utterance of the VoiceBank-2023 corpus has a WAVE file, a raw text file, and a TextGrid file. In the following, we illustrate the content of TextGrid along with WAVE file opened by Praat \cite{praat} and the metadata derived from the corpus.

\subsection{TextGrid Labeling}

\begin{figure}[htbp]
\centerline{\includegraphics[width=1.0\linewidth]{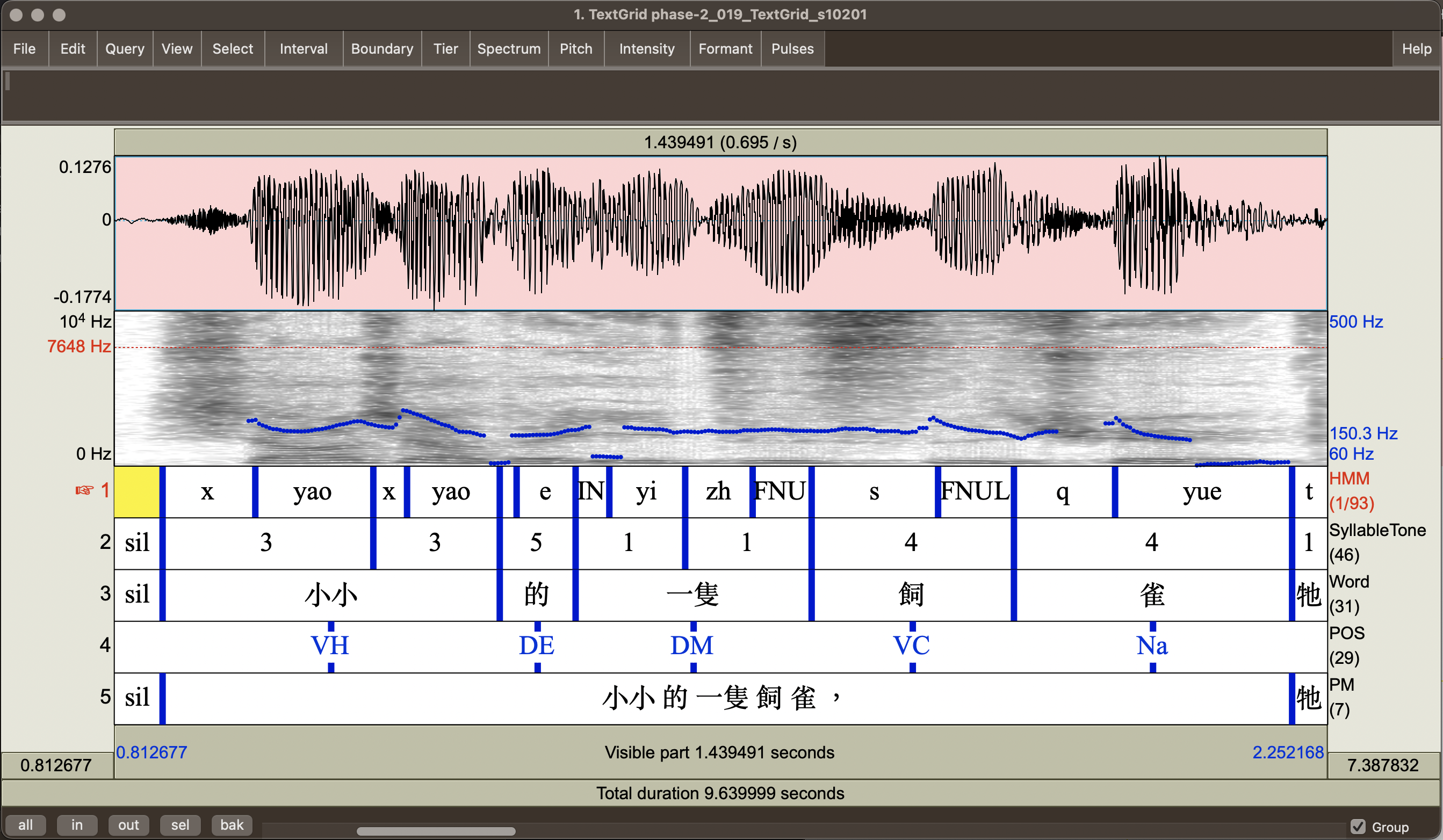}}
\caption{Examplar WAVE and TextGrid files opened by Praat.}
\label{fig:FIG4}
\end{figure}

Fig.~\ref{fig:FIG4} shows an exemplar TextGrid file opened with the corresponding WAVE file. The upper two panes are displayed waveform and spectrogram with F0 trajectory. The lower five panes show five tiers of time alignments, from top to down, HMM (acoustic units of initial and final), SyllableTone (syllable tone), Word (lexical word), POS (part of speech associated with Word), and PM (sentence-like unit delimited by punctuation marks). The TextGrid labelings of the corpus can be used in constructing speech models that need information about linguistic units and the associated time alignments.

\subsection{Metadata}
Since Phases 1-4 recorded speaks' utterances with registration information stored in a database management system, the following metadata can be obtained along with the TextGrid labeling files: speaker types, gender, the number of recording takes, time elapsed for recording each utterance (seconds), speech rate, articulation rate, estimated signal-to-noise ratio (SNR). The degree of dysarthria was labeled for each speaker with four levels (degrees). The four degrees regard the following features with informal subjective tests of listening to one to three utterances: speech intelligibility, dysarthria, fluency regarding prosody, and speaking rate. The four degrees and the corresponding characteristics are 1) degree 1: fluent speech without speech impairment, 2) degree 2: disfluent in prosody, 3) degree 3: light dysarthria but high speech intelligibility, 4) degree 4: dysarthria and low speech intelligibility. We also label each speaker with sound quality regarding recording using informal listening tests. The sound quality is measured by a five-point mean opinion score (MOS), in which 1 and 5 points represent low and high sound qualities, respectively.

Some significant statistics can be found in the metadata:
\begin{enumerate}[leftmargin=*]
    \item Speakers with higher dysarthria degrees have lower speech and articulation rates.
    \item More recording takes were made for Part-1: VoiceBanking than Part-2: Common Phrases (1.9 v.s. 1.3 takes on average), showing that short utterances alleviate the workload.
    \item Speakers' self-service recording with quality USB microphones in Phases 2-4 generally made recorded utterances have similar or even higher SNR than the on-site recorded utterances in Phase 1. This partially confirms the feasibility of online web-based recording for voice-banking.
\end{enumerate}

\section{Evaluation of Corpus for Constructing Personalized TTS Systems}
We evaluated the TTS constructed by the Part-1 VoiceBanking sub-corpus contributed by Phase-1 speakers. The TTS here is a modularized system expressed by $speech=TTS(text)=WG(SG(PG(TA(text))))$. The modules (functions) are:

\noindent1) TA: \textbf{\textit{T}}ext \textbf{\textit{A}}nalysis that generates linguistic features, and is powered by cascading the sub-modules of the rule-based text normalization \cite{txtnorm}, the CRF-based word tokenizer and part-of-speech tagger \cite{crftok}, hybrid lexicon-data driven G2P \cite{pd}.

\noindent2) PG: \textbf{\textit{P}}rosody \textbf{\textit{G}}eneration powered by the speaker adaptive prosodic model \cite{pg1} \cite{pg2} that produces prosody parameters from the linguistic features by TA.

\noindent3) SG: \textbf{\textit{S}}peech parameter \textbf{\textit{G}}eneration, which generates frame-based Mel-generalized cepstrum, logF0, and unvoiced/voiced (U/V) flag from features from PG. Note that the sub-modules of SG are similar to the conventional HTS speech synthesizer with state duration, logF0 and MGC modules. The state duration module is powered by a five-layer CNN  with speaker embedding as biases in the hidden layers. The acoustic module is powered by a stack of 4-layer CNN and 1-layer LSTM with speaker embedding as biases in the hidden layers for generating frame-based MGC, logF0, and U/V.

\noindent4) WG: \textbf{\textit{W}}aveform \textbf{\textit{G}}eneration by the WORLD vocoder \cite{world}.

The offline SLMTK 1.0 \cite{slmtk} labeled the speech corpora with linguistic-speech alignments and prosodic tags. The PG and SG were constructed with the labeling, respectively, in an adaptive and multi-speaker manner. The 15 enrolled ALS patients and 17 patients' caregivers rated the constructed personalized TTS systems with mean opinion scores for speaker similarity of 3.92 and 3.55, respectively \cite{VoiceBankProject}. The results inferred that the Part-1 VoiceBanking sub-corpus could be used to construct acceptable personalized TTS systems.

\section{Conclusions and Future Works}

This paper reports the design, recording, data purging, and correction for the VoiceBank-2023 speech corpus. The usefulness of the Part-1 (VoiceBanking) sub-corpus of VoiceBank-2023 for constructing personalized TTS systems was evaluated with reasonable MOS in speaker similarity. With the statistics of the number of recording takes and the feedback from the enrolled patients and voice donors,  we will shorten the prompts of Part-1 (VoiceBanking) sub-corpus displayed on the GUI to make speakers easier to read. Besides continuing to improve the performance of the personalized TTS systems, constructing automatic mechanisms to label the degrees of dysarthria, voice quality, and sound quality labelers will be worthwhile.

\section*{Acknowledgment}

The authors thank the 61 voice donors for banking their voices in the Asia-Pacific Medical Students' Symposium (APMSS) 2022, hosted by the NTU College of Medicine, Taiwan. The authors also thank Prof. Jing-Yi, Jeng, NKNU, Taiwan, for providing Mandarin common phrases as the prompts on the VoiceBanking website, Prof. Sin-Horng Chen and Prof. Yih-Ru Wang of NYCU, Taiwan, for providing Treebank-SR corpus and the word tokenizer and POS tagger.

\end{document}